\documentclass{cmslatex}
\usepackage{latexsym, amssymb, enumerate, amsmath}

\renewcommand{\theequation}{\arabic{section}.\arabic{equation}}

\newtheorem{thm}{Theorem}[section]

\newtheorem{lem}[thm]{Lemma}

\newtheorem{rem}[thm]{Remark}

\newcommand{\eps}{\varepsilon}

\newcommand{\be}{\begin{eqnarray}}
\newcommand{\ee}{\end{eqnarray}}
\newcommand{\ben}{\begin{eqnarray*}}
\newcommand{\een}{\end{eqnarray*}}

\newcommand{\na}{\nabla}

\begin{document}

\title{An Energetic Variational Approach  for ion transport
\thanks{
}}
\author{ Shixin Xu
\thanks {Department of Mathematics, University of Science and Technology of China,
Hefei 230026, China, (xsxztr@mail.ustc.edu.cn).}
\and Ping Sheng\thanks {Department of Physics, Hong Kong University of Science and Technology
Clear Water Bay, Kowloon, Hong Kong, China, (sheng@ust.hk).}
\and Chun Liu\thanks {Department of Mathematics, Pennsylvania State University, University Park, PA, 16802, USA (liu@psu.edu).}
}



\pagestyle{myheadings} \markboth{An Energetic Variational Approach  for ion  transport}{S. Xu P. Sheng C. Liu  }\maketitle

\begin{abstract}
The  transport and distribution of charged particles  are    crucial in the study of many physical and biological
problems. In this paper, we employ an  Energy Variational Approach to derive the coupled Poisson-Nernst-Planck-Navier-Stokes system.  All physics is included in the choices of  corresponding
energy law and kinematic transport of particles. The variational derivations give the coupled force balance equations in a unique and deterministic
 fashion. We also discuss the situations with  different types of  boundary conditions. Finally, we show that the Onsager's relation holds for the electrokinetics, near the initial time of a step function applied field.
\end{abstract}

\begin{keywords}
\smallskip Energetic Variational Approach, Poisson-Nernst-Planck (PNP) system, (Least) Action Principle, (Maximum) Dissipation Principle, Onsager's relation.

{\bf subject classifications. 35Q35, 35Q92, 76W05, 92B05.}
\end{keywords}
\section{Introduction and Background }
The Poisson-Nernst-Planck (PNP) system    
 is one of the most extensively studied  models for the   transport of charged particles in  many physical and biological
 problems,  such as free moving electrons in  semiconductors \cite{Jerome1995, Markowich1986,Markowich1990},    fuel cell \cite{Nazarov,Promislow2001},  ion particles in the  electrokinetic fluids \cite{Ben2002,Hunter2001,Jin2007,Lyklema1995}, and ion channels in cell membranes \cite{Bazant2004,Eisenberg1996,Nonner1999}.  Traditionally, the PNP system can be derived by explicit averaging of correlated Brownian trajectories \cite{Schuss2001}, while the actual dynamics of charged particles  in water and protein channels  are much more complicated  \cite{Eisenberg2006}. In  continuum description,   the PNP system can also be  viewed as the consequence of both conservation  of ion distributions  and  the Fick's law.
The limitation of this method is   that the specific interactions of  particles  are usually ambiguous or totally neglected.  The purpose of  this paper is to present an alternative way, an   Energetic Variational Approach (EnVarA) \cite{EisenbergHyon2010}, in which a consistent, coupled system of equations can be derived for the description of charged particles transport. Our approach is motivated by the seminal work of Lars Onsager  \cite{Onsager1931a,Onsager1931b}, that has an attribution to Lord Rayleigh's 1873 paper \cite{Strutt1873}.

The general framework  of EnVarA is the combination of the statistical physics and nonlinear thermodynamics.
The First Law of Thermodynamics states that the rate of change of the sum of  the kinetic energy $\mathcal{K}$ and  the
internal energy $\mathcal{U}$ is equal to the sum of the rates of change of work $\mathcal{W}$ and heat $\mathcal{Q}$, so
 $\frac {d(\mathcal{K}+\mathcal{U})}{dt}=\frac{d\mathcal{W}}{dt}+\frac{d\mathcal{Q}}{dt}.$
 From the standard statistical physics, the internal energy $\mathcal{U}$ takes into account   the particles interactions. Such interactions can be local, such as hard core interactions and nonlocal, such as Coulomb electro static interactions.
The Second Law of Thermodynamics, in the isothermal case,  is given by,
 $T\frac{d\mathcal{S}}{dt}=\frac{d\mathcal{Q}}{dt}+\Delta,$
 where $T$ is temperature, $\mathcal{S}$ is entropy and $\Delta\geq0$ is entropy production.  As a reformulation of the linear response assumption, this entropy production functional can be represented as the sum of various  rates such as the velocities and the strain rates.  By subtracting the Second Law from the First Law, under the isothermal assumption, we have,
  \be
 \frac{d}{dt}(\mathcal{K}+\mathcal{U}-T\mathcal{S})=\frac{d\mathcal{W}}{dt}-\Delta,
 \ee
where $\mathcal{F} := \mathcal{U} -\mathcal{TS}$ is  the Helmholtz free energy, and $\mathcal{K} +\mathcal{F}$ is  the total
energy $E^{total}$. In   case  no external forces or fields are applied, i.e.,  $\frac{d\mathcal{W}}{dt} = 0$, the above
expression yields the usual energy dissipation law \cite{EisenbergHyon2010,Forster2013,QianWang2006,Ryham2006}, where the entropy production  is the sole contribution to the dissipation,
\be\label{energydisspationlaw}
 \frac d {dt}E^{total}+\Delta=0\ \Leftrightarrow\ \frac d {dt}E^{total}=-\Delta.
 \ee

The (Least) Action Principle (LAP) states that  the equation of motion for a
Hamiltonian system is the direct result of  the variation of the action functional  $A=\int_0^{t^*}\!\!\!\int_{\Omega} (\mathcal{K}-\mathcal{F})dxdt$ with respect
to the flow map $x(t) = x(X, t)$ (with $x(X,0)=X$)  \cite{Arnold1989}. In other words, LAP optimizes the action with
respect to all trajectories $x(t) = x(X, t)$ by taking the variation with respect to $x$,
$\delta A=\int_0^{t^*}\!\!\!\!\int_{\Omega_0}[F_{con}]\cdot \delta xdXdt,$
where $F_{con}$ is  the conservative force and $\Omega_0$ is the Lagrangian reference domain of $\Omega$. In particular, in equilibrium,
 we have the condition $F_{con}=0$ for  a Hamiltonian dynamics.

Next, we treat the  dissipation part with  the (Maximum) Dissipation Principle (MDP) \cite{Onsager1931a,Onsager1931b,Sheng2008,HyonKwak2010}. Take the  variation with respect to the velocity (rate) in Eulerian coordinates
$\delta(\frac 1 2\Delta)=\int_{\Omega}[F_{dis}]\cdot\delta udx,$ where $F_{dis}$ is dissipative force.
  Note that the factor $\frac 1 2$ corresponds to the underlying  assumption that $\Delta$ is quadratic in the function $\mathbf{u}$. In particular, $F_{dis}$ is linear in $\mathbf{u}$, indicating the fact that  we can view  MDP as  just  a reformulation  of the linear response assumption of the nonequilibrium thermodynamics \cite{Kubo}. Such postulations are the key to  Onsager's approach \cite{Onsager1931a,Onsager1931b}, as realized by Kubo \cite{Kubo} in the more explicit linear response theory.

The final  equation of motion, the balance of all forces, includes both conservative and dissipative components.

The following auxiliary Lemma is   crucial  in  the energetic variational derivation of the system of coupled equations.
\begin{lem}\label{lnvariational}
Let f   satisfy the mass conservation law $f_t+\na\cdot(\mathbf{u}f)=0$. Define  $W=\int_{\Omega}\omega(f)$dx and  $\Pi(\omega)=\omega_ff-\omega$ , and then $ \delta W=\int_{\Omega}\na\Pi\cdot \delta x dx$.
\end{lem}

\noindent \textbf{Proof:} The conservation of mass  is equivalent to  $f(x(X,t),t)=\frac{f_0(X)}{J}$, which is a direct consequence of the identity $F_t+\mathbf{u}\cdot\na F=\na\mathbf{u} F$, where $f_0(X)$ is the initial density, $J=\det F$, $ F(X,t)=\frac{\partial x(X,t)}{\partial X}$ is   the deformation gradient  tensor \cite{Forster2013}. Rewrite the integration in the Lagrangian coordinate system and obtain
$\displaystyle W=\int_{\Omega_0}\omega(\frac{f_0(X)}{J})J dX$.
Taking the variation with respect to flow map $x\mapsto x+\eps y$, it yields
\be
\delta W&=&\frac d {d\eps} \Big{|}_{\eps=0}W(x+\eps y)  = \frac d {d\eps}\Big{|}_{\eps=0} \int_{\Omega_0}\omega(\frac{f_0(X)}{J(x+\eps y)})J(x+\eps y)dX\nonumber\\
&=&-\int_{\Omega_0}\omega_{f}(\frac{f_0(X)}{J}) \frac{f_0(X)}{J^2}\cdot tr(\frac{\partial X}{\partial x}\frac{\partial y}{\partial X})\cdot J ^2 dX+\int_{\Omega_0} \omega(\frac{f_0(X)}{J})\cdot J\cdot tr(\frac{\partial X}{\partial x}\frac{\partial y}{\partial X})dX\nonumber\\
&=&\int_{\Omega}-(\omega_{f}f-\omega,\na_x \tilde{y})dx
=\int_{\Omega}(\na(\omega_ff-\omega),\tilde{y})dx,
\ee
where $ \tilde{y}(x(X,t),t)=y(X,t)$. Hence the result holds.  $\blacksquare$
\begin{rem}\label{lnrem}
\noindent The above Lemma relates the pressure (the equation of states) to the free energy density. For given energy dissipation law $\frac d {dt} E^{total}=\frac {d}{dt}\int_{\Omega}\omega(f)dx=-\int_{\Omega}f|\mathbf{u}|^2dx$, if $\omega(f)=cf\ln f$ only contains the Gibbs entropy, i.e. no particle interactions, hence ideal gas,   then  $\Pi=\omega_ff-\omega=cf$. In particular, $f$ satisfies $f_t=c\triangle f$,  which is   a  simple diffusion equation \cite{EisenbergHyon2010}.

 If  $\omega(f)=a f^{\gamma}$, then $ \Pi= \omega_{f}f-\omega =a(\gamma-1) f^{\gamma}$. In particular, $f$ satisfies $f_t=\triangle(a(\gamma-1)f^{\gamma})$, which gives the  diffusion equation in porous media \cite{vazquez2012}.
\end{rem}

Next, we use the compressible Navier-Stokes (NS) equations as an example to illustrate the framework of EnVarA.  We start with the kinematic mass conservation,
\be
\rho_t+\na\cdot(\rho\mathbf{u})=0,
\ee
where $\rho$ is mass density of fluid and $\mathbf{u}$ is velocity of fluid. This is equivalent to the relation $\displaystyle \rho(x(X,t),t)=\frac{\rho_0(X)}{J}$, where $J=\det F $,   $ F(X,t)=\frac{\partial x(X,t)}{\partial X}$ is   the deformation gradient tensor  and $\rho_0(X)$ is the initial density \cite{Forster2013}. The following  energy dissipation law includes all the physics for these Barotropic fluids \cite{Shames1962}.
\be
\frac{d}{dt}\int_{\Omega}\left(\frac 1 2 \rho|\mathbf{u}|^2+\omega(\rho)\right)dx=-\int_{\Omega}\left[\mu_1|\na \mathbf{u}|^2+\mu_2|\na\cdot\mathbf{u}|\right]dx,
\ee
where $\mu_1$ and $\mu_2$ are viscosity constants and $\omega(\rho)$ is the Hemholtz free energy density.

By LAP and   Lemma \ref{lnvariational}, we obtain the conservative force
\be\label{Euler}
F_{con}=-\left(\rho(\mathbf{u}_t+\mathbf{u}\cdot\na\mathbf{u})+\na \Pi(\rho)\right),
\ee
with $\Pi(\rho)=\omega_{\rho}\rho-\omega$ being the pressure. By MDP, the dissipative force is  ,
\be\label{stokes}
F_{dis}=-\left(\na\cdot(\mu_1\na \mathbf{u})+\na(\mu_2\na\cdot\mathbf{u})\right).
\ee
Finally, the total force balance gives the Navier-Stokes equation,
\be
\rho(\mathbf{u}_t+\mathbf{u}\cdot\na\mathbf{u})+\na \Pi(\rho)=\na\cdot(\mu_1\na \mathbf{u})+\na(\mu_2\na\cdot\mathbf{u}).
\ee
The conservative force corresponds to the compressible  Euler equation, while the dissipative force corresponds to the Stokes equation.
  Navier-Stokes equation can be  viewed as  a hybrid model combining these two independent system.

In this paper, we use the EnVarA to derive the  electrokinetic  systems by considering the particles interactions in the dissipation part and the corresponding energy law. The outline of paper is as follows: in  $\S2$, we present the derivation of  the electrokinetic system, Poisson-Nernst-Planck-Navier-Stokes(PNP-NS) system, by using the EnVarA; in $\S3$ we focus on the energy law of the PNP-NS system with different types of boundary conditions; the Onsager relation  is proved in $\S4$; conclusion part is given in $\S5$.
\section{Derivation of Electrokinetic System Using EnVarA}
Ion transport in solutions by nature  is a multiscale-multiphysics system. With the macroscopic hydrodynamics description, the microscopic dynamics takes account of diffusion and convection as well as electrostatics. The cross scale coupling can be modeled in the general EnVarA framework. The total energy include all the equilibrium physics included in system
\be\label{pnpengergy}
E^{total}=\int_{\Omega}\underbrace{\frac {\rho}{2}|\mathbf{u}|^2dx}_{macroscopic}+ \underbrace{\left[ K_BT(n\ln \frac {n} {n_{\infty}}+p\ln \frac {p}{p_{\infty}})+\frac {\eps}{2}|\na\phi|^2\right]}_{microscopic}dx,
\ee
where $\rho$ is the mass density of fluid, $\mathbf{u}$ is the macroscopic  velocity of fluid,  $K_B$ is the Boltzmann constant,  $T$ is the absolute temperature, $n_{\infty}(p_{\infty})$ is the characteristic  negative(positive) charge distribution, $n(p)$ is negative (positive) charge distribution, the dielectrics of solution is chosen to be the  constant $\eps$, and $\phi$ is electric potential. The first term is the macroscopic kinetic energy of the solution fluids. The second and
third terms are the thermo-fluctuations (Gibbs entropy) of the ion species. The last term is the electro energy.

In the macroscopic scale, we consider the fluid to be incompressible, i.e. $\na\cdot \mathbf{u}=0$.
 At the same time, we observe the following kinematic     conservation of charge distributions,
\be\label{distribution}
n_t+\na\cdot(n\mathbf{u}_n)=0,\ \ \ p_t+\na\cdot(p\mathbf{u}_p)=0,\label{massn}
\ee
 where $\mathbf{u}_n$ and  $\mathbf{u}_p$ are  the effective velocities of negative and positive charges, respectively.
 The Gauss's law yields the Poisson equation,
\be\label{poisson}
-\eps\triangle\phi=ze(p-n),
\ee
where   $z$ is valence of ion and  $e$ is the charge for one electron. Equivalently, the potential $\phi$ can be given by the Green's kernel $G(x,y)$ in the form of
\be\label{green}
\phi(x)=ze\frac 1 {\eps} \int_{\Omega}G(x,y)(n-p)(y)dy.
\ee
By substituting  \eqref{green} into \eqref{pnpengergy}, the
energy can be written in the following form
 \be E^{total} &=&\int_{\Omega}\frac {\rho}{2}|\mathbf{u}|^2dx+
\int_{\Omega}K_BT(n\ln \frac {n} {n_{\infty}}+p\ln \frac
{p}{p_{\infty}})dx\\&&+\frac {ze}
{2\eps}\int_{\Omega}(p-n)(x)\int_{\Omega}G(x,y)(n-p)(y)dydx,\nonumber
\ee
where  the last term, the electrostatic  energy,     represents the nonlocal Coulomb interactions.

In order to take into account the more detailed  interactions of particles, we furthermore consider the dissipation functional $\Delta$   as a sum of three parts, which are all quadratic in terms of the 'rates', the velocities,
\be\label{dissipation}
\Delta=\int_{\Omega} \left [\frac {K_BT} {D_n}n|\mathbf{u}_n-\mathbf{u}|^2+\frac {K_BT} {D_p}p|\mathbf{u}_p-\mathbf{u}|^2+\eta|\na \mathbf{u}|^2\right]dx,
\ee
where $D_n \ (D_p)$ is the diffusion constant of negative (positive) ion  and $\eta$ is the viscosity of fluid. The first and second terms represent the frictions between particles and the solvents. The last term is the fraction caused by the viscosity of the solutions.


Now we begin to use the EnVarA to derive the electrokinetic system. In this case, there are  three flow maps corresponding to three velocities fields, $\mathbf{u}$, $\mathbf{u_n}$, $\mathbf{u_p}$: macroscopic flow map $x(X,t)$, negative charge map $x_n(X,t)$ and positive charge map $x_p(X,t)$, respectively. For map $x_n$,    Lemma \ref{lnvariational}, Remark \ref{lnrem} and the variation yield,
\be\label{nconforce}
F_{n-con}&=&\frac{\delta A}{\delta x_n}
 = \frac{\delta}{\delta x_n}\left[\int_0^{t^*}\!\!\!\left(\int_{\Omega}\frac {\rho}{2}|\mathbf{u}|^2dx-
\int_{\Omega}K_BT(n\ln \frac {n} {n_{\infty}}+p\ln \frac
{p}{p_{\infty}})dx\right.\right.\nonumber\\
&&-\left.\left.\frac {ze}
{2\eps}\int_{\Omega}(p-n)(x)\int_{\Omega}G(x,y)(n-p)(y)dydx\right)dt\right]\nonumber\\
&=&-(K_BT\na n-zen\na\phi)=-n\na\mu_n,
\ee
where $ \mu_n:=\frac{\delta}{\delta n} E^{total}=K_BT(1+\ln n)-K_BT\ln n_{\infty}-ze\phi$ is the chemical potential for negative charge distribution $n(x,t)$.

Using MDP, we calculate the variation of $\frac 1 2 \Delta$ with respect to the velocity $\mathbf{u}_n$ to get the dissipative force,
\be\label{ndisforce}
F_{n-dis}=\frac{\delta}{\delta \mathbf{u}_n}(\frac 1 2 \Delta)=\frac {K_B T} {D_n} n(\mathbf{u}_n-\mathbf{u}).
\ee
The  total  force balance for  negative charge yields (including \eqref{nconforce} and \eqref{ndisforce}):
\be\label{un}
n\mathbf{u}_n=n\mathbf{u}-\frac{D_n}{K_BT}n\na\mu_n.
\ee
Substituting \eqref{un} into \eqref{massn}, the mass conservation of negative charge is:
\be\label{nequation}
n_t+\na\cdot(\mathbf{u}n)=\na\cdot\left(D_n \na n-\frac{ze}{K_BT}D_nn\na\phi\right).
\ee
Similarly for positive charge, we can get
\be
p\mathbf{u}_p=p\mathbf{u}-\frac{D_p}{K_BT}p\na\mu_p,\label{up}\\
p_t+\na\cdot(\mathbf{u}p)=\na\cdot\left(D_p\na p+\frac{ze}{K_BT}D_pp\na\phi\right) \label{pequation},
\ee
where $\mu_p:=\frac{\delta }{\delta p}E^{total}=K_BT(1+\ln p)-K_BT\ln p_{\infty}+ze\phi$
is the chemical potential for positive charge distribution $p(x,t)$. In the absence of  the flow field $\mathbf{u}$,    equations \eqref{nequation} \eqref{pequation} with the Poisson equation \eqref{poisson} give the PNP system .

As for the  macroscopic flow map $x(X,t)$, considering the incompressible condition, we use 1-parameter family of volume preserving diffeomorphisms   to perform the variation, i.e.   function $x^{\eps}$ such that $x^{0}=x$, and  $\frac{dx^{\eps}}{d\eps}\Big{|}_{\eps=0}=y,$ and for any $\eps:\ \det\frac{\partial x^{\eps}}{\partial X}=1$,
which in fact leads to a divergence free condition for $y(X,t)=\tilde{y}(x(X,t),t)$, i.e. $\na_x\cdot\tilde{y}=0$.
 For LAP, we use the variations $x^{\eps}$ of x as described above  and with y satisfying $y(X,0)=y(X,t^*)=0$ for any $X\in\Omega_0$. We can calculate the variation of action functional:
\be\label{fcon1}
&&\frac{d}{d\eps}\Big{|} _{\eps=0}A(x^{\eps})=\frac{d}{d\eps}\Big{|}_{\eps=0}\int_0^{t^*}\!\!\!\!\int_{\Omega_0}\frac 1 2 \rho_0(X)|x_t^{\eps}|dXdt\\
&=&\int_{0}^{t*}\!\!\!\!\int_{\Omega_0}-\rho_0(X)(x_t)_t\cdot ydXdt=\int_0^{t^*}\!\!\!\!\int_{\Omega}-\rho(x,t)(\mathbf{u}_t+\mathbf{u}\cdot\na_x\mathbf{u})\cdot\tilde{y}dxdt.\nonumber
\ee
Hence by the Weyl's decomposition or Helmholtz's decomposition, for some $\Pi_1\in W^{1,2}(\Omega)$, we have,
\be\label{fcon}
-\rho(x,t)(\mathbf{u}_t+\mathbf{u}\cdot\na_x\mathbf{u})=\na_x\Pi_1.
\ee
By MDP and incompressible constrain, we obtain the following equation of motion for the dissipative part,
\be
  -\eta\triangle \mathbf{u}+\frac {K_BT}{D_n}n(\mathbf{u}-\mathbf{u}_n)+\frac{K_BT}{D_p}p(\mathbf{u}-\mathbf{u}_p)=\na\tilde{\Pi}_2 ,
 \ee
where $\tilde{\Pi}_2$ is the Lagrange multiplier of incompressible constrain.

Substitute   \eqref{un} and  \eqref{up} into above formula and let $\Pi_2=\tilde{\Pi}_2-K_BT\na n-K_BT\na p$,
 \be\label{fdis}
\na\Pi_2 &=& -\eta\triangle \mathbf{u} -(n-p)ze\na\phi.
 \ee
Then using the force balance,  \eqref{fcon} and \eqref{fdis} yield
\be\label{NS}
\rho(\frac{\partial\mathbf{ u}}{\partial t}+(\mathbf{u}\cdot\na)\mathbf{u})= \eta\triangle \mathbf{u}-\na\Pi+(n-p)ze\na\phi,
\ee
with $\Pi=\Pi_1-\Pi_2$. The last term is the Lorentz force induced by the charges in the fluids. It is the reaction to the convected term in \eqref{nequation} and \eqref{pequation}, which is  consistent with the Newton's third Law.
Combining \eqref{poisson}, \eqref{nequation}, \eqref{pequation}, \eqref{NS} and incompressibility, we get the coupled Poisson-Nernst-Planck-Navier-Stokes (PNP-NS) system:
\be\left\{\begin{array}{l}\label{pnps}
n_t+\na\cdot(\mathbf{u}n)=\na\cdot\left(D_n\na n-\frac{ze}{K_BT}D_nn\na\phi  \right)=-\na\cdot \mathbf{J}_n, \\
p_t+\na\cdot(\mathbf{u}p)=\na\cdot\left(D_p\na p+\frac{ze}{K_BT}D_pp\na\phi  \right)=-\na\cdot \mathbf{J}_p,\\
-\eps\triangle \phi=ze(p-n), \\
\rho(\frac{\partial \mathbf{u}}{\partial t}+(\mathbf{u}\cdot\na)\mathbf{u})=\eta\triangle\mathbf{u}-\na\Pi +(n-p)ze\na\phi,\\
\na\cdot \mathbf{u}=0.
\end{array}\right.\ee
Finally in this section, we  verify the following theorem satisfied by the derived  coupled PNP-NS system \eqref{pnps}.
\begin{thm}
With the isothermal assumption and vanishing boundary conditions, the system \eqref{pnps} satisfies the following energy dissipation law,
 \be
  \frac {d}{dt} E^{total}&=&\frac d {dt}\left[\int_{\Omega}\left(\frac {\rho}{2}|\mathbf{u}|^2 +
 K_BT(n\ln \frac {n} {n_{\infty}}+p\ln \frac
{p}{p_{\infty}})+\frac {\eps}{2}|\na\phi|^2\right)dx\right]\nonumber\\
&=&-\int_{\Omega} \left [\frac {D_n}{K_BT} n|\na\mu_n|^2+\frac {D_p}{K_BT} p|\na\mu_p|^2+\eta|\na u|^2\right]dx\nonumber\\
&=&-\int_{\Omega} \left [\frac {K_BT} {D_n}n|\mathbf{u}_n-\mathbf{u}|^2+\frac {K_BT} {D_p}p|\mathbf{u}_p-\mathbf{u}|^2+\eta|\na \mathbf{u}|^2\right]dx\nonumber\\
&=&-\Delta.\ee
 Reversely, if we choose the action functional as
$$ A=\int_0^{t^*}\!\!\!\int_{\Omega}\frac {\rho}{2}|\mathbf{u}|^2dx-\int_{\Omega}K_B T(n\ln \frac {n} {n_{\infty}}+p\ln \frac
{p}{p_{\infty}})dx-\frac {\eps}{2}|\na\phi|^2 dx,$$
and the dissipation functional as  \eqref{dissipation}, then by (Least) Action Principle and (Maximum) Dissipation Principle, under the kinematic  assumption of distribution (conservation law) \eqref{distribution} and Poisson equation, we can obtain the Poisson-Nernst-Planck-Navier-Stokes system  \eqref{pnps}.
  \end{thm}

 Sketch of Proof:   From the energy law to PNP-NS system are above derivations.
  By adding the first equation multiplied  by $\mu_n$, second equation  multiplied by $\mu_p$, and fourth  equation  multiplied by $u$ together, and using the weak form of Poisson equation, we can get the energy law. $\blacksquare$

\begin{rem}
Some more complicated models can be derived by including more coupling terms for particle interactions in the total energy  $E^{total}$, such as
\begin{enumerate}
  \item  In \cite{EisenbergHyon2010,HyonFonseca2012}, it is  shown that, by EnVarA,  a   modified model can be derived naturally for ion particles with finite size effects, through adding interaction term  $E^{repulsion}=\sum_{i=1}^{N}\sum_{j\geq i}^N\frac 1 2\int_{\Omega}\Psi_{i,j}(|x-y|)c_i(x)c_j(y)dxdy$ to the total energy, where  $c_i$, $c_j$ mean different species of ions, and $\Psi_{i,j}(|x-y|)=\frac{\eps_{ij}(a_j+a_i)}{|x-y|^{12}}$ is Lennard-Jones (LJ) potential for ith and jth ions located at x and y with the radii $a_i$, $a_j$ , respectively.
  \item If we add $E^{surface}=ze\int_{\Omega}(p-n)\Psi_s$ to the total energy, we can derive the surface potential trap model \cite{WanXu} to describe the electrokinetics induced by  the interface of solid and solution,
  where $\Psi_s$ is a surface potential   only depending  on the property of material.
\end{enumerate}
\end{rem}

\section{ Boundary Conditions}
In electrokinetics, most physically interesting properties arise from different boundary conditions \cite{Chapman,LeeLee2011,Xu2011}. These boundary conditions represent the interactions between particles in the bulk solutions and  the particles in  or near  the boundary \cite{WanXu}. The  interactions can also be included into the  energy functionals.
As in the previous  sections, we assume  the non-flux boundary condition $\mathbf{J}_n\cdot\mathbf{\nu}=\mathbf{J}_p\cdot\mathbf{\nu}=0$   for charge density, with $\mathbf{\nu}$ being out normal vector, and the nonslip boundary condition  $\mathbf{u}=\mathbf{0}$ for velocity.
 We will focus on the boundary effect of potential $\phi$ which plays an important role in  electrodynamics.  For the three different boundary conditions, the PNP-NS system has the following theorem.
 \begin{thm}\label{pnpenergylaw}
  If $n$, $p$  satisfy  $\mathbf{J}_n\cdot\mathbf{\nu}=\mathbf{J}_p\cdot\mathbf{\nu}=0$, and $\mathbf{u}=\mathbf{0}$ on the boundary $\partial\Omega$, then
 \begin{enumerate}
   \item if $\phi=\phi_0(x)$, i.e. Dirichlet boundary, then PNPNS satisfies the energy law,
\be
\frac{d}{dt}E^{total} &=&\frac{d}{dt}\left[ \int_{\Omega}\frac {\rho} 2|\mathbf{u}|^2+ K_BT(n\ln \frac {n} {n_{\infty}}+p\ln \frac {p}{p_{\infty}})+\frac {\eps}{2}|\na\phi|^2\right]dx\\
&=&-\left[\int_{\Omega}\frac{D_n}{K_BT}n|\na\mu_n|^2+\frac{D_p}{K_BT}|\na\mu_p|^2+\eta|\na \mathbf{u}|^2dx\right]+\eps\int_{\partial\Omega}\frac{\partial\phi}{\partial\nu}\phi_0dx;\nonumber
\ee
   \item if $\frac{\partial\phi}{\partial \nu}=\frac{\sigma_0(x)}{\eps}$, i.e. Neumann boundary condition, then PNPNS satisfies the energy law,
   \be
\frac{d}{dt}E^{total} &=&\frac{d}{dt}\left[ \int_{\Omega}\frac {\rho} 2|\mathbf{u}|^2+K_BT(n\ln \frac {n} {n_{\infty}}+p\ln \frac {p}{p_{\infty}})+\frac {\eps}{2}|\na\phi|^2\right]dx\\
&=&-\left[\int_{\Omega}\frac{D_n}{K_BT}|\na\mu_n|^2+ \frac{D_p}{K_BT}|\na\mu_p|^2+\eta|\na \mathbf{u}|^2dx\right]+\int_{\partial\Omega}\sigma_0\phi dx;\nonumber
\ee
   \item   if $\phi+\zeta\frac{\partial\phi}{\partial n}=\phi_0(x)$, i.e Robin boundary condition,  then PNPNS satisfies the energy law,
\be\label{robin}
\frac{d}{dt}E^{total} &=&\frac{d}{dt}\left[ \int_{\Omega}\frac {\rho} 2|\mathbf{u}|^2+K_BT(n\ln \frac {n} {n_{\infty}}+p\ln \frac {p}{p_{\infty}})+\frac {\eps}{2}|\na\phi|^2+\frac{\eps}{2\zeta}\int_{\partial\Omega}|\phi|^2dx\right]\nonumber \\
&=&-\left[\int_{\Omega}\frac{D_n}{K_BT}|\na\mu_n|^2+\frac{D_p}{K_BT}|\na\mu_p|^2+\eta|\na \mathbf{u}|^2dx\right].
\ee
 \end{enumerate}
 \end{thm}
\begin{rem}
\begin{enumerate}
  \item When $\phi$ on the boundary is Robin boundary condition, as time approaches infinity,  \eqref{robin} means $\na\mu_n=\na\mu_p=\na \mathbf{u}=\mathbf{0}$.
Considering the boundary condition, it yields $\mathbf{J}_n=\mathbf{J}_p=\mathbf{u}=\mathbf{0}$, which means there is no fluid flux in the time limit. Then we can derive a Charge Conservation Poisson Boltmann (CCPB) equation \cite{LeeLee2011,WanXu}
\be
-\eps\triangle\phi=zen_{\infty}V\left(\frac{\beta\exp(-ze\phi/K_BT)}{\int_{\Omega}\exp(-ze\phi/K_BT)dx}-\frac{\alpha\exp(ze\phi/K_BT)}{\int_{\Omega}\exp(ze\phi/K_BT)dx}\right)\nonumber
\ee
 as the time limit of PNPNS system, where $\alpha=\frac{n_0}{n_{\infty}}$ and $\beta=\frac{p_0}{n_{\infty}}$ with $n_0 ~(p_0)$ being the initial negative (positive) ion distribution.

  \item In the ion transport process, most of time an extra field is added to the domain to generate the electrodynamics phenomena. When there is
an external filed added to the PNPNS system, there will be an extra term $\int_{\Omega}(p-n)\Psi dx$ added to the total energy in Theorem \ref{pnpenergylaw}, where $-\na\Psi$ is the extra electric field \cite{Jackson1999,WanXu}.
\end{enumerate}
\end{rem}
\section{Onsager's Relation in Cylindrical Situation}
The coupling between the flow field and the electric field  gives arise to all the important properties and applications of the electrokinetic fluids.
For instance, when the fluid-solid interface is charged,  the application of an electrical voltage difference
can   induce a fluid flow. This effect is known as electroosmosis
(EO). Conversely, the application of a pressure gradient can
generate, besides fluid flow, a voltage difference that is denoted
as the streaming potential (SP). The EO and SP coefficients
are not independent. They are related by the well-known Onsager's reciprocal relation \cite{Onsager1931a,Onsager1931b}.
It  dictates that the electric current density $\mathbf{J}_e$ and the fluid current density $\mathbf{J}_f$  be linearly related to the voltage gradient $\na\phi$  and the pressure gradient $\na\Pi$ :
\be\label{onsager}\left[\begin{array}{l}\mathbf{J}_e\\
                         \mathbf{J}_f\end{array}\right]=-\left[\begin{array}{ll} L_{11}&L_{12}\\
                                                                      L_{21}& L_{22}\end{array}\right]\left[\begin{array}{l}\na\phi\\
                         \na\Pi \end{array}\right],
 \ee                  		
where  $L_{11}$ is the electrical conductivity and  $L_{21}$ is the hydrodynamic permeability. In literature \cite{Xu2011}, the proportional matrix is treated as symmetric and attributed to Onsager's  relation. Onsager's  reciprocal relation, the microscopic reversibility \cite{Onsager1931a,Onsager1931b} is a stability conditions. It is manifested by specific coupling effects in different physical settings. In \eqref{onsager}, it is  a reformulation of the fact that Lorentz force and the transport of charge are action  and  reaction.

 Onsager's reciprocal relation has many forms in different settings. Here  we take  the axisymmetric cylinder coordinate for low Reynolds number situations with
  constant initial values, i.e. $p(\cdot,0)=p_0,~n(\cdot,0)=n_0$  as an example. Then the  PNP-NS system  is simplified to be  the Poisson-Nernst-Planck-Stokes (PNP-S) system.
  If an extra filed $E_z$  and a pressure drop $\frac{\partial\pi}{\partial z}$ are  added in $z$ direction,
 the velocity $u_z$ satisfies
\be
 \frac{\partial \Pi}{\partial z}-\mu\left[\frac 1 r\frac{\partial }{\partial r}(r\frac{\partial u_z}{\partial r})\right] =(p-n)ze(-\frac{\partial\phi}{\partial z}+E_z),
\ee
with $u_z(r=a)=\phi(r=a)=0$ .
 At the initial several steps, $\frac{\partial \phi}{\partial z}$, $\frac{\partial p}{\partial z}$, $\frac{\partial n}{\partial z}$ and $\frac{\partial u_z}{\partial z}$  are small and negligible.
Then we can get
\be
u_z=\frac{\eps E_z\phi}{\mu} +  \frac{a^2-r^2}{4\mu}(-\frac{\partial\Pi}{\partial z}).
\ee
And the fluxes for the negative and positive charges in z direction are
 \be J_n&=&-\left(D_n\frac{\partial n}{\partial z}-D_n\frac{ze}{K_BT}(\frac{\partial\phi}{\partial z}-E_z)n\right)=-\frac{ze}{K_BT}nD_nE_z\\
 J_p&=&-\left(D_p\frac{\partial p}{\partial z}+D_p\frac{ze}{K_BT}(\frac{\partial\phi}{\partial z}-E_z)p\right)=\frac{ze}{K_BT}pD_pE_z.
\ee
The total electric  current  in z direction is the sum of the current carried (transported) by the flow field $\mathbf{u}$ and the current due to the electric field,
\be
J_e&=&\frac{\int_{-L}^L\int_0^a2(p-n)zeu_zrdrdz+\int_{-L}^L\int_0^a 2ze(J_p-J_n)rdrdz}{2La^2}\nonumber\\
&=&E_z\left[\frac{\eps }{\mu a^2}\int_0^a rze(p-n)dr+\frac{z^2e^2}{K_BT} E_z(D_pp_0+D_nn_0)\right]\nonumber\\
&&+(-\frac{\partial \Pi}{\partial z})\frac{1}{2\mu a^2}\int_{0}^a(p-n)ze(a^2-r^2)rdr.
\ee
The fluid  flux in z direction is
\be
J_f&=&\frac{\int_{-L}^{L}\int_0^a2u_zrdr}{2La^2}=\frac{2\eps E_z}{\mu a^2}\int_0^a \phi rdr+(-\frac{\partial\Pi}{\partial z})\frac{1}{2\mu a^2}\int_0^a(a^2-r^2)rdr\nonumber.
\ee
We may write the Onsager relation as
\be\left[\begin{array}{l}J_e\\
                         J_f\end{array}\right]=\left[\begin{array}{ll} L_{11}&L_{12}\\
                                                                      L_{21}& L_{22}\end{array}\right]\left[\begin{array}{l}E_z\\
                         -\frac{\partial \Pi}{\partial z} \end{array}\right],
 \ee
where we have introduced the function forms of the coefficients,
\be
L_{12}=\frac{1}{2\mu a^2}\int_0^a rze(p-n)(a^2-r^2)dr,~~
L_{21}=\frac{2\eps}{\mu a ^2}\int_0^a\phi rdr.
\ee
But by the Poisson equation, $L_{12}$ can be rewritten as
\be
L_{12}&=& - \frac {\eps}{2\mu a^2}\int_0^a(a^2-r^2)(\frac 1 r\frac{\partial\phi}{\partial r}+\frac{\partial^2\phi}{\partial r^2})rdr\nonumber\\
&=&-\frac{\eps}{2\mu a^2}\left[(-a^2\phi(0)+2\int_{0}^a\phi rdr)+(a^2\phi(0)-\int_0^a 6r\phi dr)\right]\nonumber\\
&=&\frac{2\eps}{\mu a^2}\int_0^ar\phi dr=L_{21},
\ee
which gives the symmetric property of the matrix.

\section{Conclusion}
In this paper, we derive the  electrokinetic system  for  ion transport in solutions by using an Energy Variational Approach.  Taking into consideration of particles interactions in both the free energy functional and the dissipation functional, we obtain the Poisson-Nernst-Planck-Navier-Stokes system. We can extend our theory to include more detailed description of the solutions, such as  the finite size effects of the charged particles and various boundary effects. Since the boundary condition of potential play an important role in electrokinetic, we also present the boundary effects to the energy law.
 The energy laws with an external electric field  under different boundary conditions of potential are also obtained.  A short demonstration of Onsager's relation is presented for
the Poisson-Nernst-Planck-Stokes system under cylinder axisymmetric coordinate.
\section{Acknowledgment }
P. S. wishes to acknowledge the support of SRFI11/SC02 and RGC grant HKUST604211 for this work.   C. L. and S. X wishes to acknowledge the partial support by the NSF grants DMS-1109107, DMS-1216938 and DMS-1159937.  S. X. also wishes to acknowledge China Scholarship Council for support.

\end{document}